\documentclass{article}[12pt]

\usepackage{amstext,amsmath,amssymb,amsfonts,bbm}
\usepackage[latin1]{inputenc}
\usepackage[OT2,T1]{fontenc}
\usepackage{epsfig}
\usepackage{dsfont}
\usepackage{hyperref}
\usepackage{amsthm}
\usepackage{color}
\usepackage{multirow}
\usepackage{psfrag}
\usepackage{graphicx}
\usepackage{extarrows}

\newtheorem{theorem}{Theorem}

\begin{document}
\author{Thomas Krajewski\footnote{The author declares that there is no conflict of interest regarding the publication of this paper.}\\
Centre de Physique Théorique\\
Aix-Marseille Université\\
{  \tt thomas.krajewski@cpt.univ-mrs.fr}\\
}
\title{A renormalisation group approach to the universality of Wigner's semicircle law for random matrices with dependent entries }
\date{5th Winter Workshop on \\Non-Perturbative Quantum Field Theory\\ Sophia-Antipolis, March 2017}
\maketitle

\abstract{In this talk, we show that if the non Gau\ss ian part of the cumulants of a random matrix model obey some scaling bounds in the size of the  matrix, then Wigner's semicircle law holds. This result is derived using the replica technique and an analogue of the renormalisation group equation for the replica effective action.
}

\section{Introduction}

Random matrix theory (see the classical text \cite{Mehta}) first appeared in physics in Wigner's work on the level spacing in large nuclei. Since then, it has proved to have multiple applications to physics and other branches of science, see for instance \cite{Oxford}.  Most of these applications rely on the universal behaviour of some of the observables for matrices of large size. A simple example is Wigner's semicircle law for the eigenvalue density  that holds in the large $N$ limit for matrices whose entries are independent and identically distributed.

Understanding the universal behaviour of eigenvalue distributions and correlations  ranks among the major problems in random matrix theory. In this respect, the renormalisation group turns out to be a powerful technique. Introduced in the context of critical phenomena in statistical mechanics by K. Wilson to account for the universality of critical exponents, the latter has also proved to be useful in understanding probability theory. For instance it leads to an insightful proof of the central limit theorem, see the review by  G. Jona-Lasinio \cite{CLT} and references therein.
 
The renormalisation group has been used to derive the semicircle law for random matrices in the pioneering work  of E. Brézin and A. Zee \cite{CR}. In the latter approach, the renormalisation group transformation consists in integrating over the last line and column of a matrix of size $N+1$ to reduce it to a size $N$  matrix. This leads to a differential equation for the resolvent $G(z)=1/N\langle \text{Tr}\,(z-M)^{-1}\rangle$  in the large $N$ limit whose solution yields the semicircle law.

In this talk, we follow a different route: We first express the resolvent as an integral over replicas and introduce a differential equation for the replica effective action. This differential equation is a very simple analogue of Polchinski's exact renormalisation group equation \cite{Polchinski}. It is used to derive inductive bounds on the various terms, ensuring that the semicircle law is obeyed provided the cumulants of the original matrix model fulfil some simple scaling bounds in the large $N$ limit.

This talk is based on some work in collaboration with A. Tanasa and D.L. Vu in which we extend Wigner's law to random matrices whose entries fail to be independent \cite{paper}, to which we refer for further details. There have been other works on such an extension, see \cite{dependent1}, \cite{dependent2} and \cite{dependent3}.

\section{What are random matrices ?}

A random matrix is a probability law on a space of matrices, usually given by the joint probability density on its entries, 
\begin{align}
\rho(M)=\rho(M_{11},M_{12},\dots)
\end{align}
Thus a random matrix of size $N$ is defined as a collection of $N^{2}$ random variables. However, there is a much richer structure than this, relying notably on the spectral properties of the matrices.

Here we restrict our attention to a single random matrix. Note that it is also possible to consider several random matrices, in which case the non commutative nature of matrix multiplication plays a fundamental role, leading to the theory of non commutative probabilities.
 
There are two important classes of probability laws on matrices.

\begin{itemize}
\item Wigner ensemble: The entries are all independent variables,
\begin{align}
\rho(M)=\prod_{i,j}\rho_{ij}(M_{ij}),
\end{align}
up to the Hermitian condition $\overline{M}_{ij}=M_{ji}$. 
\item  Unitary ensemble: The probability law is invariant under unitary transformations  
\begin{align}
\rho(UMU^{\dagger})=\rho(M),
\end{align}
for any unitary matrix $U\in\text{U}(N)$. 
\end{itemize}

The only probability laws that belong  to  both classes are the Gau\ss ian ones
\begin{align}
 \rho(M)\propto \exp-{\textstyle \frac{1}{2\sigma^{2}}}\text{Tr}(M)^{2},
\end{align}
up to a shift of $M$ by a fixed scalar matrix.

The main objects of interest are the expectation values of observables, defined as
\begin{align}
\langle{\cal O}\rangle=\int  dM\rho(M) {\cal O}(M).
\end{align}
Among the observables, the spectral observables defined as symmetric functions of the eigenvalues of $M$, play a crucial role in many applications. This is essentially due to their universal behaviour: In the large $N$ limit, for some matrix ensembles and in particular regimes, the expectation values of specific spectral observables do not depend on the details of the probability law $\rho(M)$. 

Universality is at the root of the numerous applications to physics and other sciences, since the results we obtain are largely model independent. Among the applications to physics, let us quote the statistics of energy levels in heavy nuclei, disordered mesocopic systems, quantum chaos, chiral  Dirac operators, ...

\section{Wigner's semicircle law}

In this talk, we focus on the eigenvalue density, defined as
\begin{align}
\rho(\lambda)=\frac{1}{N}\Bigg\langle \sum_{1\leq i\leq N}\delta\bigg[\lambda-\lambda_{i}\bigg(\frac{M}{\sqrt{N}}\bigg)\bigg]\Bigg\rangle.
\end{align}
In particular, a universal behaviour is expected in the large $N$ limit for some ensembles. 

\begin{figure}[h]

\begin{center}
\parbox{6cm}{\includegraphics[width=6cm]{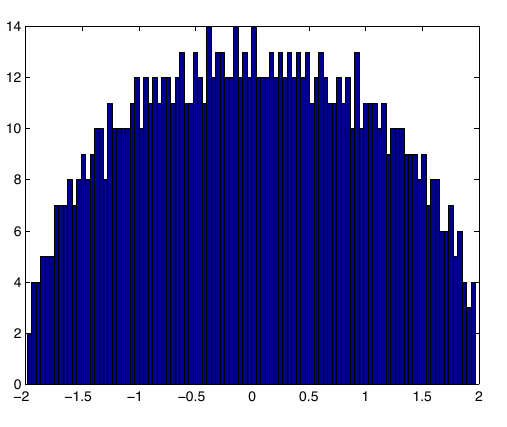}}

\caption{Spectrum of a large ($N=1000$) Hermitian matrix with a Gau\ss ian distribution ($\sigma=1$)}
\label{histogram}

\end{center}
\end{figure}

For a Gau\ss ian random Hermitian matrix $ {\displaystyle \rho(M)\propto \exp-{\textstyle \frac{1}{2\sigma^{2}}}\text{Tr}(M^{2})}$,  the eigenvalue density obeys Wigner semicircle law,
\begin{equation}
   \lim_{N\rightarrow\infty} \int_{\mathbb R} d\lambda \,\lambda^{k}\rho(\lambda)
   =\begin{cases}
   \frac{1}{2\pi \sigma^{k+2}}\int_{-2\sigma}^{2\sigma} d\lambda\,\lambda^{k}\sqrt{4\sigma^{2}-\lambda^{2}}&\text{if $k$ is even},\\
0&\text{if $k$ is odd}.\label{semicircle:eq}
\end{cases}
\end{equation}
Empirically, $\rho(\lambda)$ may be determined by plotting the histogram of eigenvalue of a matrix taken at random with a given probability law, see figure \ref{histogram}.

The derivation of Wigner's semicircle in the large $N$ limit is based on the resolvent (also known as the Green function)
\begin{align}
G(z)=\lim_{N\rightarrow+\infty}
\frac{1}{N}\bigg\langle\text{Tr}\Big(z-\frac{M}{\sqrt{N}}\Big)^{-1}\bigg\rangle
=\frac{z}{2\sigma^{2}}\bigg(1-\sqrt{1-\frac{4\sigma^{2}}{z^{2}}}\bigg).
\end{align}
Then, the density of eigenvalues is recovered as
\begin{align}
\rho(\lambda)=\frac{G(\lambda-\mathrm{i}0^{+})-G(\lambda+\mathrm{i}0^{+})}{2\mathrm{i}\pi},
\end{align}
where we have used the relation 
\begin{align}
\frac{1}{x\pm\mathrm{i}0^{+}}=\text{p.v.}\frac{1}{x}\mp 2\mathrm{i}\pi\delta(x).
\end{align}
In the large $N$ limit, for the Gau\ss ian model, the resolvent obeys the self consistency equation (also known as the Schwinger-Dyson equation), see for instance \cite{Zee}, section VII.4,
\begin{equation}
    G(z)=\sum_{k=0}^{\infty} \frac{\sigma^{k}G^{k}(z)}{z^{k+1}}=\frac{1}{z-\sigma  G(z)}.
\end{equation}
 Its solution that behave as $1/z$ for large $z$ is
\begin{equation}
    G(z)=\frac{z}{2\sigma^{2}}\bigg(1-\sqrt{1-\frac{4\sigma^{2}}{z^{2}}}\bigg).\label{Gsm}
\end{equation}
Taking the cut of the square root on the negative real axis, we obtain the Wigner semicircle law \eqref{semicircle:eq} in the large $N$ limit.

The semicircle law is not limited to the Gau\ss ian case, it also holds for Wigner matrices in the large $N$ limit.  A random Hermitian $N\times N$ matrix is a Wigner matrix if 

\begin{itemize}
\item real and imaginary parts of upper diagonal elements are independent and identically distributed (i.i.d.) with mean 0 and variance $\sigma$;
\item diagonal elements are i.i.d. with finite mean and variance and independent of the off diagonal ones.
\end{itemize}

Then, in the limit $N\rightarrow+\infty$, the eigenvalue distribution of $\frac{M}{\sqrt{N}}$ is the semicircle law \eqref{semicircle:eq}.

The original proof is of combinatorial nature and involves the expectation of the moments
\begin{align}
\lim_{N\rightarrow+\infty}\frac{1}{N^{k/2+1}}\big\langle\text{Tr}(M^{k})\big\rangle=
\begin{cases}
 \frac{(2l)!}{(l!)^{2}(l+1)}&\text{for $k=2l$ even}\\
 0&\text{for $k$ odd}.
 \end{cases}
\end{align}
To derive this result, the idea is to first factorise $\rho$ for a Wigner ensemble as
\begin{align}
\rho(M)=\prod_{i}\rho'(M_{ii})\quad\prod_{i<j}\rho''(\text{Re }M_{ij})\rho''(\text{Im }M_{ij}),
\end{align}
where $\rho'$ is the common probability density of the real diagonal terms and $\rho''$ the common probability density  of the real and imaginary parts of the off diagonal terms.

Then, we expand the trace and integrate over the independent real variables $M_{ii}$, $\text{Re}\,M_{ij}$ and $\text{Im}\,M_{ij}$. The power of $N$ in the expectation of a given moment arises from the denominator $\frac{1}{N^{k/2+1}}$ and from the number of independent indices in the summations. In the large $N$ limit, the only configurations that survive are counted by Catalan numbers, $C_{l}=\frac{(2l)!}{(l!)^{2}(l+1)}$. Since the latter also appear in the following Taylor expansion
\begin{align} 
\frac{z}{2\sigma^{2}}\bigg(1-\sqrt{1-\frac{4\sigma^{2}}{z^{2}}}\bigg)=\sum_{l\geq 0}\frac{(2l)!}{(l!)^{2}(l+1)}\frac{\sigma^{2l}}{z^{2l+1}},
\end{align}
we conclude that
\begin{align}
G(z)&=\lim_{N\rightarrow+\infty}\frac{1}{N}\Big\langle\text{Tr}\Big(z-\frac{M}{\sqrt{N}}\Big)^{-1}\Big\rangle\\
&=\lim_{N\rightarrow+\infty}\sum_{k=0}^{\infty} \frac{1}{z^{k+1}}\frac{1}{N^{k/2+1}}\big\langle\text{Tr}(M^{k})\big\rangle\\
&=\frac{z}{2\sigma^{2}}\bigg(1-\sqrt{1-\frac{4\sigma^{2}}{z^{2}}}\bigg).
\end{align}
This is the form of the resolvent that leads to Wigner's semicircle law. Here, we see universality at work: In the large $N$ limit, the eigenvalue density is given by the semicircle law, whatever the probability densities $\rho'$ and $\rho''$ are. However, this result relies on the independence of the matrix elements. In the next section, we will extend it to matrices whose entries are not necessary independent.

\section{Wigner's law beyond Wigner ensembles}

Let us introduce the cumulants, defined through their generating function
\begin{equation}
\langle M_{i_{1}j_{1}}\cdots M_{i_{k}j_{k}}\rangle_{\text{c}}=
\frac{\partial}{\partial J_{j_{1}i_{1}}}\dots\frac{\partial}{\partial J_{j_{k}i_{k}}}\log\big\langle\exp \text{Tr}(MJ)\big\rangle\Big|_{J=0}.
\end{equation}
In the physics terminology, these are the connected correlation functions. In particular, the Gau\ss ian  cumulants vanish beyond the quadratic term 
\begin{align}
 \rho(M)\propto \exp-{\textstyle \frac{1}{2\sigma^{2}}}\text{Tr}(M^{2})\Rightarrow
 \begin{cases}\langle M_{ij} M_{kl}\rangle_{\text{c}}=
\sigma^{2}\delta_{il}\delta_{jk}\\ \text{vanish otherwise}\end{cases}.
\end{align}
Therefore, cumulants of degree higher than 2 are a measure of the deviation from the Gau\ss ian case.

\begin{figure}[h]
\begin{equation*}
\begin{array}{cc}
\includegraphics[width=4cm]{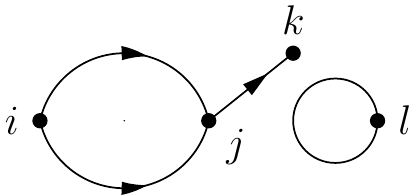}\qquad&\qquad\includegraphics[width=2cm]{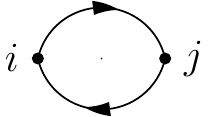}\\
\langle (M_{ij})^{2}M_{jk}M_{ll}\rangle_{\text{c}}\qquad&\qquad\langle M_{ij}M_{ji}\rangle_{\text{c}}
\end{array}
\end{equation*}
\caption{Examples of graph associated to cumulants}
\label{graphexamples}
\end{figure}

Turning back to the general case, for each cumulant we construct an oriented graph as follows (see figure \ref{graphexamples} for some examples):

\begin{itemize}
\item vertices are distinct matrix indices in the cumulant,
\item there is an edge from $i$ to $j$ for every $M_{ij}$.
\end{itemize}

Since non quadratic cumulants measure deviations from the Gau\ss ian case, if the perturbation is small it is reasonable to expect that the semicircle law is still obeyed. 

To state this result, recall that an oriented graph is Eulerian if every vertex has an equal number of incoming and outgoing edges. Equivalently, it means that every connected component admits an Eulerian cycle, i.e. an oriented cycle that passes through all edges, respecting the orientation. Furthermore, let us denote  $v(G)$, $e(G)$, $c(G)$  the number of vertices, edges and connected components of $G$.

\begin{theorem}[Wigner's law for matrices with dependent entries]
\label{Wignerdependent:thm}
Let $\rho_{N}$ be a probability law on the space of Hermitian $N\times N$ matrices $M$ such that its cumulants can be decomposed as 
$C_{G}=C_{G}^{'}+C_{G}^{''}$, with $C_{G}^{'}$ a Gaussian cumulant and $C_{G}^{''}$ a perturbation such that, uniformly in the vertex indices $i_{1},..., i_{v(G)}$ (i.e. all constants involved should not depend on these indices),
\begin{itemize}
   \item ${\displaystyle \lim_{N\rightarrow\infty}\, N^{v(G)-c(G)-e(G)/2}C''_{G}(i_{1},..., i_{v(G)})=0}$ if $G$  is Eulerian,
    \item ${\displaystyle N^{v(G)-c(G)-e(G)/2}C''_{G}(i_{1},..., i_{v(G)})}$ bounded if $G$ is not Eulerian.
\end{itemize}
Then, the moments of the eigenvalue distribution of the matrix $\frac{M}{\sqrt{N}}$ converge towards the moments of the semicircle law, with $\sigma$ given by the Gaussian cumulant  $\langle M_{ij} M_{kl}\rangle_{\text{c}}=\sigma^{2}\delta_{il} \delta_{jk}$,
\begin{equation}
   \lim_{N\rightarrow\infty} \int_{\mathbb R} d\lambda \,\lambda^{k}\rho_{N}(\lambda)
   =\begin{cases}
   \frac{1}{2\pi \sigma^{k+2}}\int_{-2\sigma}^{2\sigma} d\lambda\,\lambda^{k}\sqrt{4\sigma^{2}-\lambda^{2}}&\text{if $k$ is even},\\
0&\text{if $k$ is odd}.
\end{cases}
\end{equation}

\end{theorem}

For instance, for the graph \parbox{2cm}{\includegraphics[width=2cm]{cumulant1.pdf}} which is not Eulerian, with $v=3$, $e=4$ and $c=2$, the cumulant should obey
\begin{align}
\frac{1}{N}\big|\langle (M_{ij})^{2}M_{jk}M_{ll}\rangle_{\text{c}}\big|\leq K,
\end{align}
with $K$ a constant that does not depend on the indices $i$, $j$, $k$ and $l$. On the other hand, for the graph \parbox{1.2cm}{\includegraphics[width=1.2cm]{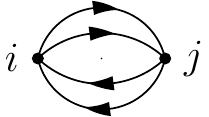}}, which is Eulerian,  with $v=2$, $e=4$ and $c=1$, we impose
\begin{align}
\lim_{N\rightarrow+\infty} \frac{1}{N}\big|\langle (M_{ij})^{2}(M_{ji})^{2}\rangle_{\text{c}}\big|=0,
\end{align}
uniformly in $i$ and $j$.

As an illustration, we recover the case of Wigner matrices (with finite moments). Indeed, 
\begin{itemize}
\item there is no graph with $v\geq 3$ (independence of off diagonal matrix elements);
\item for $v=1$ and $v=2,e\geq 3$, bounds are satisfied because of $1/N^{e/2}$ and all moments are assumed to be finite;

\item $C_{\includegraphics[width=1cm]{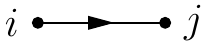}}(i,j)=\langle M_{ij}\rangle_{c}=\langle M_{ij}\rangle=0$
        (off diagonal elements have mean value 0);
       
\item $C_{\includegraphics[width=1cm]{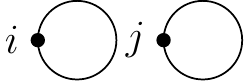}}(i,j)=\langle M_{ii}M_{jj}\rangle_{c}=
       \langle M_{ii}M_{jj}\rangle-
       \langle M_{ii}\rangle\langle M_{jj}\rangle=0$ (independence of diagonal elements);
       \item $C_{\includegraphics[width=1.1cm]{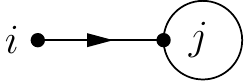}}(i,j)=\langle M_{ij}M_{jj}\rangle_{c}=
       \langle M_{ij}M_{jj}\rangle-
       \langle M_{ij}\rangle\langle M_{jj}\rangle=0$ (independence of diagonal and off diagonal elements);

              \item    $C_{\includegraphics[width=1cm]{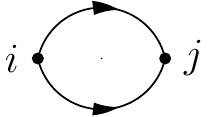}}(i,j)=\langle M_{ij}M_{ij}\rangle_{c}=
       \langle M_{ij}M_{ij}\rangle-
       \langle M_{ij}\rangle\langle M_{ij}\rangle$\\$
    \qquad\qquad \qquad\qquad\quad=  \langle (\text{Re} M_{ij})^{2}-(\text{Im} M_{ij})^{2}\rangle+2\text{i}
       \langle \text{Re} M_{ij}\text{Im} M_{ij}\rangle=0$\\
        (independence of real and imaginary parts and equality of their distributions with mean value 0);

        \item  $C_{\includegraphics[width=1cm]{cumulant2}}=\sigma^{2}$ is the Gau\ss ian cumulant leading to the semicircle law.
\end{itemize}

The case of unitarily invariant matrices is critical since the bounds are saturated, see \cite{paper}. This is consistent since we known the semicircle law is not obeyed by unitary non Gau\ss ian ensembles \cite{BIPZ}.

It is possible to give a combinatorial proof  of this result based on the relation between moments and cumulants,  
\begin{equation}
\langle M_{i_{1}j_{1}}\cdots M_{i_{k}j_{k}}\rangle=
\sum_{I_{1},\dots,I_{p} \text{ partition of} \atop
\left\{(i_{1},j_{1}),\cdots,(i_{k},j_{k})\right\}}
\big\langle\prod_{ij\in I_{1}} M_{ij}\big\rangle_{\text{c}}\cdots
\big\langle\prod_{ij\in I_{p}} M_{ij}\big\rangle_{\text{c}}\label{momentsandcumulants:eq}
\end{equation}
In the moment method, we have to estimate 
\begin{equation}
\frac{1}{N^{k/2+1}}\big\langle\text{Tr}(M^{k})\big\rangle=\frac{1}{N^{k/2+1}}
\sum_{1\leq i_{1},\dots,i_{k}\leq N}\langle M_{i_{1}i_{2}}\cdots M_{i_{k}i_{1}}\rangle.\label{trace:eq}
\end{equation}
Then, we express the moments in \eqref{trace:eq} in terms of cumulants using \eqref{momentsandcumulants:eq} and represent each cumulant as a graph. Because of the trace, one has to draw Eulerian cycles on the graphs after some vertex identifications. Then, the scaling bounds on the cumulants can be used to show that only Gaussian terms survive.

\section{Proof based on the replica effective action}

Let us give a renormalisation group proof of this result based on the replica effective action. The use of replicas in random matrix theory is a classical subject, see for instance \cite{replica-book} or \cite{replica}. To begin with, let us note that
\begin{align}
\text{Tr}\bigg(z-\frac{M}{\sqrt{N}}\bigg)^{-1}=\frac{\partial}{\partial z}\log\det\bigg(z-\frac{M}{\sqrt{N}}\bigg).
\end{align}
It is convenient to express the logarithm using the replica method. First observe that
\begin{align}
\log(A)=\mathop{\lim}\limits_{n\rightarrow 0} \frac{A^{n}-1}{n}.
\end{align}
Then, we express the $n^{\text{th}}$ power of the determinant as a Gau\ss ian integral over $n$ replicas of a complex vector of size $N$ (with a factor of $\pi^{nN}$ included in the measure),
\begin{align}
\frac{1}{\det^{n}(z-M)}= \int dX\,
  \exp-\text{Tr}\big(X^{\dagger}(z-M)X\big),
\end{align} 
which fit into a $N\times n$ complex matrix $X=(X_{i,a})_{1\leq i\leq N\atop 1\leq a\leq n}$. 

The limit $n\rightarrow 0$ may be worrisome, its meaning is as follows. Because of $\text{U}(n)$ invariance, any perturbative result in powers of $1/z$ is a polynomial in $n$, from which we retain only the linear term. Of course, this may not hold beyond perturbation theory, where replica symmetry breaking can occur.
  
After averaging over $M$ with the random matrix density $\rho(M)$, we obtain the following expression for the resolvent,
 \begin{equation}
   G(z)=-\frac{1}{N} \frac{\partial}{\partial z}
   \bigg[\!
   \int dX\,
  \exp\left\{-\text{Tr}(X^{\dagger}X)+V_{0}(X)\right\}
   \bigg]_{\text{order 1}\atop\text{ in $n$}},\label{Greplica:eq}
   \end{equation}
  where the replica potential is
\begin{align}
V_{0}(X)=\log\Big\langle\exp\text{Tr}\Big(X^{\dagger}\frac{M}{\sqrt{N}}X\Big)\Big\rangle.
\end{align} 
Because of the logarithm, the potential involves the cumulants and can be expanded over graphs as
\begin{multline}
V_{0}(X)=\\
\sum_{G\atop \text{ oriented graph}}\hskip-0.3cm{\textstyle \frac{1}{|\text{Aut}(G)|N^{e(G)/2}}}\sum_{1\leq i_{1},\dots,i_{v(G)}\leq N\atop\text{all different}}C_{G}(i_{1},\dots, i_{v(G)})\prod_{e\text{ edge}}(XX^{\dagger})_{i_{s(e)}
i_{t(e)}}\label{effectiveexpansion:eq},
\end{multline}
where $s(e)$ is the source of edge $e$ and $t(e)$ its target.

 Let us introduce a replica effective action, obtained by a partial integration
 \begin{equation}
    V(t,X)=\log\int d Y\,
  \exp\left\{-\frac{\text{Tr}( Y^{\dagger} Y)}{t}+V_{0}( X+ Y)\right\}-Nn\log\,t.\label{effective:eq}
\end{equation}
The parameter $t$ ranges between 0 (where we have no integration, $V(t=0, X)=V_{0}(X)$) and $t=1/z$.
 
\begin{figure}[h]
\begin{equation*}
\frac{\partial}{\partial t}\,\parbox{1.1cm}{\includegraphics[width=1.1cm]{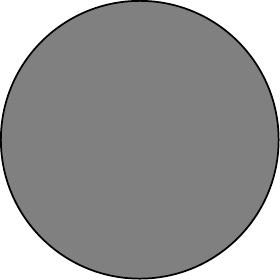}}
\qquad=\qquad
\parbox{1.7cm}{\includegraphics[width=1.7cm]{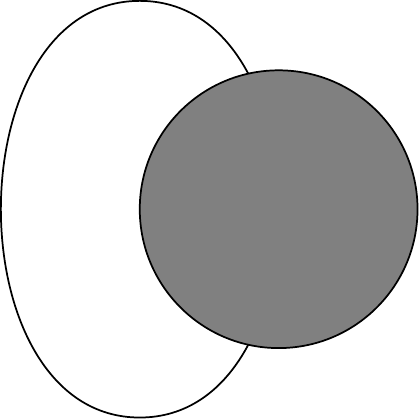}}
\qquad+\qquad
\parbox{2.3cm}{\includegraphics[width=2.3cm]{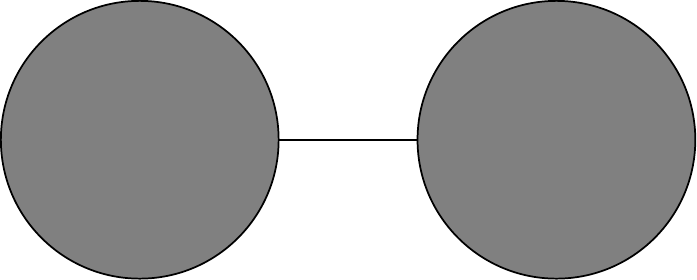}}
\end{equation*}
\caption{Graphical interpretation of the renormalisation group equation}
\label{Polch:fig}

\end{figure}

The effective potential obeys a semi-group property that follows from Gau\ss ian convolution (see for instance \cite{Zinn}, section {A10.1}),
\begin{equation}
    V(t+s, X)=\log\int d Y\,
  \exp\left\{-\frac{\text{Tr}(Y^{\dagger} Y)}{s}+V( t,X+Y)\right\}-Nn\log\, s.
\end{equation}
For small $s=dt$, it translates into the following renormalisation group equation, which is a simple version of Polchinski's exact renormalisation group equation \cite{Polchinski}, 
\begin{equation}
    \frac{\partial V(t,X)}{\partial t}=
    \sum_{i,a}
    \Bigg(
    \frac{\partial^{2}V(t, X)}{\partial X_{i,a}\partial\overline{ X}_{i,a}}+
    \frac{\partial V(t, X)}{\partial X_{i,a}}
    \frac{\partial V(t,X)}{\partial\overline{ X}_{i,a}}
    \Bigg).\label{RGdiff:eq}
    \end{equation}
The first term on the RHS is referred to a the loop term, since it creates a  new loop in the Feynman graph expansion of the effective action while the second inserts a one particle reducible line and is referred as the tree term, see figure \ref{Polch:fig}.

Taking into account the boundary condition $V(t=0, X)=V_{0}(X)$, it is convenient to write \eqref{RGdiff:eq} in integral form
\begin{equation}
    V(t,X)=V_{0}(X)+\int_{0}^{t}ds\,
    \sum_{i,a}
    \Bigg(
    \frac{\partial^{2}V(s, X)}{\partial X_{i,a}\partial\overline{ X}_{i,a}}+
    \frac{\partial V(s, X)}{\partial X_{i,a}}
    \frac{\partial V(s,X)}{\partial\overline{ X}_{i,a}}
    \Bigg).\label{RG:eq}
\end{equation}
This allows us to derive inductive bounds in powers of $t=1/z$.

From a physical point of view, we evaluate the effective potential by a large succession of small partial integrations, with a total weight given by $t$. Let us stress that in our context this differential equation is merely a tool to control the $t$ dependence of the effective action after integrating with a $t$ dependent propagator.

\begin{figure}[h]
\begin{equation*}
\begin{array}{cccc}
\begin{minipage}{2cm}
${\displaystyle\frac{\partial^{2}V}{\partial X_{i,a}\partial\overline{ X}_{i,a}}}$\\
\small (loop term)
\end{minipage}
\,:\quad&\parbox{2cm}{\includegraphics[width=2cm]{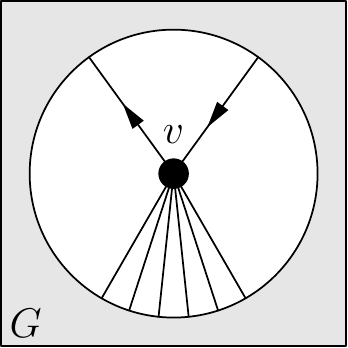}}
&\quad\rightarrow\quad&
\parbox{2cm}{\includegraphics[width=2cm]{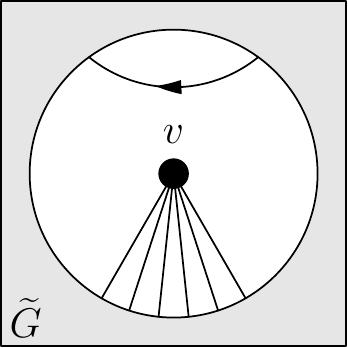}}\\
\\
\begin{minipage}{2cm}
${\displaystyle\frac{\partial V}{\partial X_{i,a}}
    \frac{\partial V}{\partial\overline{ X}_{i,a}}}$\\
     \small (tree term)

    \end{minipage}
    \,:\quad
    &\parbox{2cm}{\includegraphics[width=2cm]{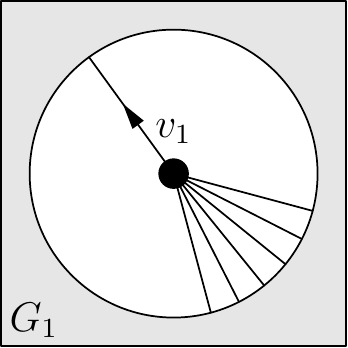}}
\quad\parbox{2cm}{\includegraphics[width=2cm]{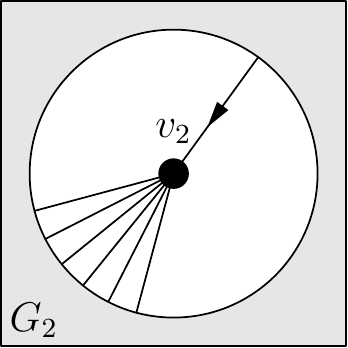}}&\quad\rightarrow\quad&
\parbox{2cm}{\includegraphics[width=2cm]{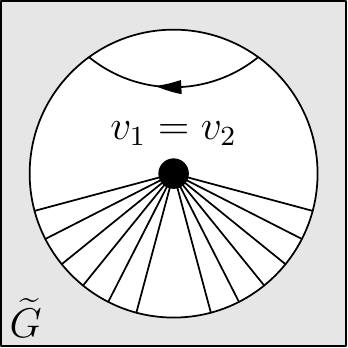}}
\end{array}
\end{equation*}
\caption{Action of the differential operators on the vertices of the effective action}
\label{RGgraphical}

\end{figure}

The effective potential also admits an expansion over graphs,  
\begin{multline}
V(t, X)=\\
\sum_{G\atop \text{ oriented graph}}\hskip-0.3cm{\textstyle \frac{1}{|\text{Aut}(G)|N^{e(G)/2}}}\sum_{1\leq i_{1},\dots,i_{v(G)}\leq N\atop\text{all different}}C_{G}(t;i_{1},\dots, i_{v(G)})\prod_{e\text{ edge}}(XX^{\dagger})_{i_{s(e)}
i_{t(e)}}\label{effectiveexpansion:eq}.
\end{multline}
This leads to a graphical interpretation of the action of the two differential operators in the renormalisation group equation, see figure \ref{RGgraphical}.  Indeed, in the expansion \eqref{effectiveexpansion:eq}, an edge joining a vertex carrying label $i$ to a vertex carrying $j$ is equipped with a factor $\sum_{a}{X}_{i,a}\overline{X}_{j,a}$, with $a$ a replica index. Then, the differential operator $\frac{\partial}{\partial {X_{i,a}}}$ (resp. $\frac{\partial}{\partial \overline{X}_{j,a}}$) removes the outgoing (resp. incoming) half edge.  Finally, the remaining half edges are reattached  and the vertices identified to yield a new graph on the RHS of $\eqref{RG:eq}$, with one less edge. These operations are performed on the same graph for the loop term and on distinct ones for the tree term.

Let us decompose the effective cumulants appearing in \eqref{effectiveexpansion:eq} into Gau\ss ian ones  and perturbations, and expand both in a power series in $t=1/z$,
\begin{align}
C_{G}(t)=\sum_{k=0}^{\infty}t^{k} \big[\underbrace{C^{'(k)}_{G}}_{\text{Gau\ss ian}}\!+\!\underbrace{C^{''(k)}_{G}}_{\text{perturbation}}\!\!\!\big].
\end{align}
The Gau\ss ian terms are those that are constructed using only the Gau\ss ian term in the initial potential  $V_{0}(X)$.  Even is $V_{0}(X)$ is quartic in $X$, this does not hold for the  Gau\ss ian part of $V_{t}(X)$, that contains terms of all orders. The perturbation collects all the remaining terms,  they contain at least one non Gau\ss ian perturbation from $V_{0}(X)$.

The renormalisation group equation \eqref{RG:eq} allows us to prove inductively on $k$  that the perturbations $C^{''(k)}_{G}$ obey the same scaling bound imposed on $C^{''(0)}_{G}=C^{''}_{G}(0)$ and that the purely Gau\ss ian terms do not grow to fast,

\begin{itemize}
   \item ${\displaystyle \lim_{N\rightarrow\infty}\, N^{v(G)-c(G)-e(G)/2}\big[C^{''(k)}_{G}\big]_{\text{order 0 }\atop\text{in $n$}}=0}$ for $G$ Eulerian,    
   \item ${\displaystyle N^{v(G)-c(G)-e(G)/2}\big[C^{''(k)}_{G}\big]_{\text{order 0 }\atop\text{in $n$}}}$ bounded for $G$ not Eulerian,
\item ${\displaystyle N^{v(G)-c(G)-e(G)/2} \big[C^{'(k)}_{G}\big]_{\text{order 0 }\atop\text{in $n$}}}$  bounded for any  $G$.
\end{itemize}

This involves a combinatorial discussion based on the graphical interpretation of figure \ref{RGgraphical} that can be found in \cite{paper}. Let us simply mention that the terms that may violate the bounds are of higher order in $n$. Thus, they are harmless when taking the limit $n\rightarrow 0$ before the limit $N\rightarrow+\infty$.

Finally, using \eqref{Greplica:eq} and the renormalisation group equation \eqref{RG:eq}, the resolvent can be expressed as 
\begin{equation}
    G(z)=\frac{1}{z}+\frac{1}{N^{3/2}z^{2}}\sum_{1\leq i\leq N}\big[C_{\includegraphics[width=0.8cm]{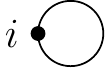}}(1/z;i) \big]_{\text{order 0 in $n$}}\label{Greenpotential:eq}.
 \end{equation}
The scaling bounds for the non Gau\ss ian cumulants impose, perturbatively in $1/z$, 
\begin{align}
\lim_{N\rightarrow\infty}\, \frac{1}{\sqrt{N}}\big[C''_{\includegraphics[width=0.8cm]{cumulantii}}(1/z;i) \big]_{\text{order 0 in $n$}}=0.
\end{align}
Therefore, only the Gau\ss ian cumulants contribute and we recover Wigner's semicircle law.

\section{Conclusion and outlook}

In this talk, we have argued that Wigner's semicircle law remains valid for matrices with dependent entries. The deviation from the independent case is measured by the joint cumulants of the entries, which are assumed to fulfil some scaling bound for large $N$. To establish this result, we have introduced an effective action for the replicas. This effective action obeys a renormalisation group equation that allowed us to prove perturbative bounds on the effective cumulants. As a consequence of these bounds, only the Gau\ss ian terms contribute in the large $N$ limit, thus establishing the validity of Wigner's semicircle law.

It may also be of interest to investigate the case of the sum of a random matrix $M$ and a deterministic one $A$, see for instance \cite{replica} where such a model is discussed. In this case, the resolvent is expressed as
 \begin{equation}
   G(z)=-\frac{1}{N} \frac{\partial}{\partial z}
   \bigg[\!
   \int dX\,
  \exp\left\{-\text{Tr}(X^{\dagger}(A+z)X)+V_{0}(X)\right\}
   \bigg]_{\text{order 1}\atop\text{ in $n$}}.
   \end{equation}
 In our context, the  deterministic matrix $A$ induces a non trivial kinetic for the replicas. In particular, if $A$ is a discrete Laplacian, it yields a non trivial renormalisation group flow that bears some similarities with the QFT renormalisation group. In this case, we expect to exploit the true power of the renormalisation group equation, with a discussion of fixed points and scaling dimensions.

\paragraph{Acknoledgements}I wish to thank the organisers of the workshop, Mario Gattobigio, Thierry  Grandou and Ralf Hoffmann  for their kind invitation as well as Stan Brodsky for insightful remarks. I am also grateful to my collaborators, Dinh Long Vu  and Adrian Tanasa.




\end{document}